\documentclass[useAMS,usenatbib]{mn2e}

\title[Bow nebulae]
{The role of wind driving in OB star bow nebulae}

\author[C. Struck]
{Curtis Struck\thanks{E-mail: curt@iastate.edu}\\
Dept. of Physics and Astronomy, Iowa State Univ., Ames, IA 50011 USA}

\usepackage{amssymb}
\usepackage{amsmath}
\usepackage{graphicx}
\usepackage{multirow}
\def\aap{{ A\&A}}

\def\aj{{AJ}}

\def\apj{{ApJ}}
\def\apjl{{ApJL}}

\def\mnras{{MNRAS}}

\def\apjs{{ApJS}}

\begin{document}
\date{\today}

\pagerange{\pageref{firstpage}--\pageref{lastpage}} \pubyear{0000}

\maketitle

\label{firstpage}
\begin{abstract}
Bow-shaped mid-infrared emission regions have been discovered in satellite observations of numerous late-type O and early-type B stars with moderate velocities relative to the ambient interstellar medium. Previously, hydrodynamical bow shock models have been used to study this emission. It appears that such models are incomplete in that they neglect kinetic effects associated with long mean free paths of stellar wind particles, and the complexity of Weibel instability fronts. Wind ions are scattered in the Weibel instability and mix with the interstellar gas. However, they do not lose their momentum and most ultimately diffuse further into the ambient gas like cosmic rays, and share their energy and momentum. Lacking other coolants, the heated gas transfers energy to interstellar dust grains, which radiate it. This process, in addition to grain photo-heating, provides the energy for the emission. A weak R-type ionization front, formed well outside the infrared emission region, generally moderates the interstellar gas flow into the emission region. The theory suggests that the infrared emission process is limited to cases of moderate stellar peculiar velocities, evidently in accord with the observations. 

\end{abstract}


\section{Introduction: Infrared Bow Nebulae}
\label{intro}

Hundreds of late-type O and early-type B stars have been discovered with crescent or bow shaped emission regions in the infrared, located near the star.  \citet{kobulnicky16} have summarized the discovery history of these `arctuate' bow nebulae associated with OB stars which generally have significant peculiar velocities and strong mid-to-far infrared emission. The first, and among the closest, of these objects were discovered by \citet{gull79} with optical emission line images. Using {\it IRAS} far-infrared observations \citet{van95} catalogued 58 candidates, though this list was subsequently re-analysed and reduced to 19 objects \citep{noriega97}. With the advent of mid-infrared data from the Spitzer and WISE space telescopes more detailed studies of small samples were carried out and ultimately larger samples were catalogued. These include the survey of several hundred runaway stars by \citet{peri12, peri15}, and the direct searches of archival IR imagery for these objects by \citet{kobulnicky16, kobulnicky17}. The latter works identified over 700 objects.

The bow nebulae are usually found within a few tenths of a parsec from their stars, which are typically moving a couple tens of km s$^{-1}$ relative to the surrounding gas. These stars generally have substantial winds, and \citet{kobulnicky18} have found that the net emission provides a good measure of the stellar mass loss. The nebulae are not generally seen in early O-types or hypervelocity stars. 

Since the discovery of $\zeta$ Ophiucus by \citet{gull79} the arctuate nebulae have been identified with hydrodynamic bow shocks produced by the wind of a moving star interacting with the ambient interstellar gas. A literature of numerical simulations of paired bow shocks around moving stars has grown up along with the observational discoveries. These models (e.g.,  \citealt{comeron98, mackey13, meyer14, mackey15, meyer16, acreman16, scherer16,  meyer17, green19, scherer19}) have been compared to a variety of different types of moving stars and to the analytic bow shock equations of \citet{wilkin96}. Hydrodynamic bow shock models for arctuate nebulae are supported by the fact that they are usually found at radii expected from estimates of the standoff radius given the wind parameters, the stellar peculiar velocity, and an estimate of the density of the surrounding interstellar medium. We note that the bow shock models have focussed primarily on higher relative velocities than are typical for the IR nebulae (e.g., $>$ 30 km s$^{-1}$), and the models of \citet{acreman16}, in particular, use a rather high ISM density. Thus, the models are most relevant to hyper-velocity stars, see Sec. 5. 

Another important characteristic of the infrared bow nebulae, long recognized, is that they are contained within the HII region of their parent star; the radius of the ionization edge is often an order of magnitude greater, see Sec. 2. It is well known (e.g., \citealt{draine11}) that in moving HII regions a weak ionization-shock forms at the edge of the ionization bubble. A number of papers have modeled this structure (e.g., \citealt{mackey13, mackey15,  green19})), but its effects on and relation to the inner infrared bow nebulae have not been widely considered. 

This and other impediments to obtaining a complete picture of these objects may be due in part to the fact that both components are rarely seen together. Indeed, images of the (faint) outer ionization-shock are hard to acquire. Exceptions include the nearby object RCW 120, described in \citet{mackey15}, and the Bubble Nebula, see Sec. 4. A wholistic model, assembled from these pieces, will be discussed in Sec. 2. 

Yet another important characteristic of bow nebulae is that the contact region between the stellar wind and the interstellar gas is likely mediated by the plasma Weibel instability rather the collisional processes that normally are responsible for the abrupt shock jumps in hydrodynamic quantities. The role of the Weibel instability has been recognized in the literature (e.g., \citealt{caprioli14b}), but not the kinetic complexity of the mixing region it implies. It is often simply assumed that sharp shocks form in this region, though, as discussed below, there is more complexity in the present case, where tenuous, high velocity winds hit a denser ionized gas. In this paper we will explore kinetic effects in the extended particle mixing zone, especially the energy deposition of high velocity wind ions (Sec. 2.2), and the role this deposition may play as a heating source for the infrared emission. .

\section{The Model}
\label{sect:model}

To develop a simple conceptual model we assume that the star is surrounded by and moving through a uniform, neutral HI medium, which contains a magnetic field of typical galactic strength. Secondly, we will consider the problem in the frame of the star, so the medium is viewed as a steady wind (the ISM wind) directed towards the star and its own wind (the star wind). In this section we will focus on the case of stars moving through the ISM with velocities of order a few tens of km s$^{-1}$, as is the case for most observed IR bow nebulae. We will also limit consideration to times when a steady flow has developed, and neglect transients or large amplitude fluctuations.

\subsection{Near the Str\"{o}mgren edge}

Late-type O stars or early B stars have classical Str\"{o}mgren radii $\sim$10-70 pc over a range of spectral types from B0.5 to O6, assuming a gas density of about $n_{H+} \simeq 1.0\ cm^{-3}$ (e.g., \citealt{draine11}, Ch. 15). These radii scale with the gas density as $r_S \sim n_{H^+}^{2/3}$. Generally, ionization edges, even in stars moving through the ISM, are quite sharp, see e.g., \citet{osterbrock89, draine11}.

This sharpness is also seen in numerical models of runaway O-stars, e.g., \citet{mackey13, mackey15}. More precisely, the velocities considered here are close to or somewhat greater than twice the sound speed in the ionized region (e.g., c = 13 km s$^{-1}$ at T = 8000K), so these are weak R-type fronts (see Draine, Ch. 24). For such fronts the post-front density is about twice the initial, and the post-front velocity about half the pre-front value in the frame of the front, and in the smooth hydrodynamic flow limit. 

Ordered magnetic fields in the ISM can be important, the models of \citet{mackey15} showed how they affect the flow around an HII bubble and compress the leading edge. Behind the ionization front, the fields will interact with all of the gas, not just a small, pre-shock ionized fraction, and contribute another source of pressure. For typical interstellar field strengths, the magnitude of this pressure may of order 10\% relative to the initial ram pressure. 

Besides ordered magnetic fields, fluctuating fields, with magnitudes comparable to those of the ordered fields, and on a range of scales below that of the HII regions are also likely \citep{spangler13}. These will also scatter particles, and add to the thermal pressure. 

The gyroradii of charged particles orbiting around the fields are very small, for the protons, 

\begin{equation}
\label{eqa}
a = \frac{mcv_{\perp}}{qB}
= 3.1 \times 10^7 \left( \frac{1.0\ \mu G}{B} \right)
\left( \frac{v_{\perp}}{30\ km\ s^{-1}} \right)\ cm,
\end{equation}

\noindent where the mass and charge of the proton have been used for $m, q$ in the second equality (see \citealt{shu92}). Variations in the magnetic field strength and directions will lead to differences in the gyromotions, and collisions between clumps of charged particles. 

\begin{figure}
\centerline{
\includegraphics[scale=0.35]{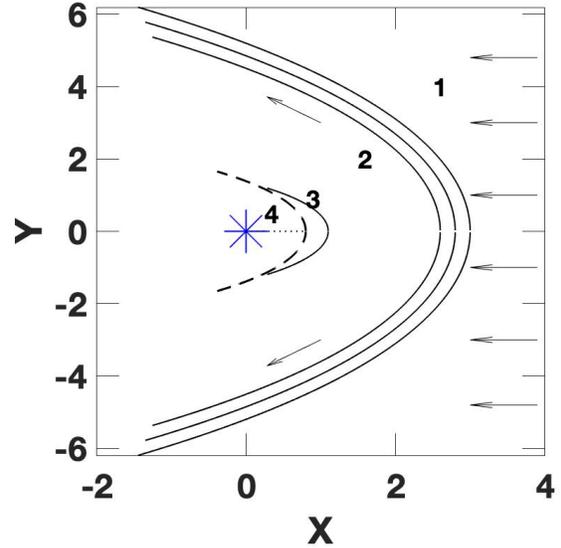}}
\caption{Schematic, not to scale, of the steady flow around OB star bow nebulae, in units of the standoff radius. Labelled region 1 is the undisturbed ISM wind as seen in the star's frame. Region 2 contains ionized hydrogen flowing trans or subsonically. The (\citet{wilkin96} shaped) curves between regions 1 and 2 represent the ionized edge, and weak J-shock. The crescent shaped region 3 is the IR emission zone, where wind and ISM particles mix. The dashed curve represents inner shock formed as a result of the Weibel instability. Region 4 consists primarily of the stellar wind.}
\label{fig:foursnap}
\end{figure}

For late-type O stars with temperatures less than 40,000 K, helium is ionized at somewhat smaller radii. However, the He - H$^+$ collisional mean free path is much less than a parsec for reasonable values of the density and velocity dispersion in this region \citep{draine11}, so the He will also be thermalized.
The post-front gas will not cool by a large amount because the photoionization heating balances cooling at a temperature of about 8000 K as in other HII regions. Only faint emissions would be expected from this front (but see possible examples discussed in Sec. 4). Behind this frontal region, there will be a roughly steady transsonic flow. 

\subsection{The mixing and IR emission region}
\subsubsection{General collision dynamics}

We turn now from the ionization front shocks at scales of order 10 pc to the infrared emission region at typical scales of a few tenths of a pc (see Fig. 1). This is also the scale where the ram pressure of the stellar and (unshocked) ISM winds would be equal, and thus, where it has been assumed that paired bow shocks form. This is an important boundary in the present model as well, where the ram and thermal pressure of the ISM flow meet the supersonic stellar wind. 

To begin, we estimate the mean free path of stellar wind protons impacting the transsonic ISM flow on these scales. For proton-proton collisions we use a variation of the argument in \citet{shu92}. First we assume that in collisions that generate large angle scattering events the electrostatic potential at closest approach at least equals the kinetic energy of the relative motion. The latter is essentially the kinetic energy of the stellar wind or beam protons, since their velocity is so high. This energy equality can be solved for the radius in the Coulomb potential, which is identified with an effective scattering radius, and then squared to obtain an effective cross section. This cross section and the ambient density can then be used to estimate the mean free path. We obtain,

\begin{equation}
\label{eqb}
\lambda_+ = \frac{m_p^2 v_{rel}^4}{n e^4}
\simeq 2.7 \left( \frac{10\ cm^{-3}}{n} \right)
\left( \frac{v_{rel}}{2000\ km\ s^{-1}} \right) ^4kpc,
\end{equation}

\noindent where $m_p$ is the proton mass, $e$ is its charge, $n$ is the ambient density, and $v_{rel}$ is the relative velocity. The second equality shows that this free path is huge, and it appears that, the stellar protons are not stopped in the ISM material, and do not make a collisional shock.

We can make the estimate of equation \eqref{eqb} for collisions between electrons in the two flows by replacing the proton mass with the electron mass. This reduces the mean free path by about a factor of $10^6$, to about 0.001 parsec. This suggests prompt thermalization between electrons in any mixing of the two flows.

The mean free path for significant scattering of ISM electrons by fast protons is also short. Each such collision will not greatly change the trajectory of a stellar proton. However, at a rate of about one electron scattering per mean free path traversed, their effects will accumulate as an effective drag. This kinetic drag is not the same as conventional dynamical friction. Except for rare very close encounters, the velocity impulses imparted to most ambient electrons by passing stellar protons are much smaller then typical thermal velocities, so a density-enhanced wake will not be produced. After of order 1000 close scattering events a proton will have scattered its mass in electrons, and lost much of its kinetic energy. This corresponds to a travel distance of about 0.5 pc, assuming an average speed of 1000 km s$^{-1}$, half the wind speed. This is an overestimate since the mean free path diminishes rapidly as the speed diminishes according to equation \eqref{eqb}. 

Thus, the electrons can take up and share wind energy a the winds decelerate. This energy is also shared with ambient ISM protons, since the mean free path for that interaction is also short. These processes are equivalent to cosmic ray heating and will be discussed further below. 

\subsubsection{Weibel turbulence zone}

Collisional particle processes aside, we expect a mixing front to form between the two flows as a result of the collisionless Weibel instability between two plasma streams (\citealt{weibel59} and also the review of \citealt{blandford87}, Sec. 6.3). In this instability, small scale, but highly turbulent, electric currents and magnetic fields are generated in filaments, which effectively scatter the particle streams. Detailed simulations are presented in \citet{caprioli14a, caprioli14b} and \citet{bohdan20}. 

 At the intersection of the two flows analytic perturbation calculations suggest a characteristic instability growth time of $\gamma \simeq \omega_p v_{rel} / c$  \citep{weibel59}. The plasma frequency is $\omega^2_p = 4{\pi}{n_w}e^2/m$, where here $n_w$ is the wind density. The corresponding length scale is $c/\omega_p$, and the length scale corresponding to the Weibel growth rate is, 

\begin{equation}
\label{eqbb}
l_g = \frac{c}{\omega_p}
\left( \frac{c}{v_o} \right)
\simeq 5.1 \times 10^8 \left( \frac{0.0020\ cm^{-3}}{n_w} \right) cm,
\end{equation}

with, 

\begin{multline}
\label{eqbc}
n_w = 0.0020 \left( \frac{\dot{M}}{3.0 \times 10^{-8} M_\odot\ yr^{-1}} \right)
\left( \frac{2000\ km\ s^{-1}}{v_{rel}} \right)\\
\times \left( \frac{0.2\ pc}{R_o} \right) ^2\ cm^{-3}.
\end{multline}

Even if the Weibel instability takes many of these characteristic lengths to grow to saturation, the overall scale is small, and the instability develops rapidly. The decay of the Weibel filaments and resulting thermalization take longer. Specifically, the models of \citet{caprioli14a, caprioli14b} indicate that it takes more than several thousands of plasma length scales. This gives a total Weibel zone scale of, for example, $3000c/\omega_p \simeq 1.5 \times 10^{12}\ cm$ or about $0.10\ a.u.$. This is very small compared to other scales considered here, so there is indeed something like a Weibel shock front at the small radius boundary of a bow nebula. 

Wind particles traveling through this Weibel zone will likely be scattered by the induced fields, as magnetic Wiebel filaments first saturate, then dissipate. The simulations of \citet{caprioli14a, caprioli14b} show that the momentum distribution of fast particles in Weibel turbulence evolves modestly, and particle momenta are not dissipated overall. The distribution broadens, and in fact, some particles are accelerated. The situation in these models is somewhat different from the present case, because the fast particles are added to the colliding flows that generate the Weibel turbulence, as when cosmic rays hit a supernova remnant shock. In the present case, the fast particles constitute one of the colliding flows. Nonetheless it is likely that a significant population of fast wind particles emerge from the most intense Wiebel turbulence into the larger bow mixing region, albeit with scattered trajectories. This makes sense because scattering by the dominantly magnetic fluctuations of Weibel turbulence will dissipate little energy. However, some energy must be lost from both wind and ISM flows to maintain the turbulence.

The work of \citet{bret15} and \citet{bret20} on particle trajectories in colliding flows also indicates that fast particles are not trapped and can escape the strong turbulence region, preferentially in directions aligned with the filaments, which tend to be predominantly along the flow direction. \citet{bret20} shows that in the case of an external magnetic field with a significant component along the flow direction, particles escape on chaotic trajectories. Randomly oriented fields in the local ISM likely satisfy these conditions frequently. 

Assuming a substantial fraction of the wind energy is scattered past the strong Weibel instability zone, it will then dissipate, e.g., via electron drag (and other processes including wave generation), in the bulk of the bow nebula, which is shared with the ambient gas. A thermal pressure gradient will be produced in this mixing zone, and the mildly supersonic ISM inflow will be slowed and turned aside. In contrast to the classic hydrodynamic models, there will be no outer shock paired with the Weibel front, unless the peculiar velocity of the star is much greater than the range considered here. 

\subsubsection{Hot gas zone}

The discussion above suggests that the kinetic energy of the wind is dissipated into thermal energy and plasma waves in the mixing region. This is a substantial heating, but is it able to overcome the recombination and (ionized metal) line cooling, which in classical HII regions balances photo-ionization heating to keep the temperature near 8000 K? Assuming a moderately high value of that cooling of $\Lambda/n_e n_p = 3 \times 10^{-24}$ ergs cm$^3$ s$^{-1}$ (see e.g., \citealt{draine11}), we can estimate the cooling luminosity of a bow nebula of volume $V$ as, 

\begin{multline}
\label{eqbd}
L_{cool} 
= 7.1 \times 10^{31}
 \left( \frac{n_e}{10\ cm^{-3}} \right)
 \left( \frac{n_p}{10\ cm^{-3}} \right)\\
\times
\left( \frac{V}{(0.2\ pc)^3} \right)\ ergs\ s^{-1},
\end{multline}

\noindent where, $n_e,\ n_p$ are the electron and proton number densities in the emission region.

The (hemispheric) wind luminosity is, 

\begin{multline}
\label{eqh}
L_w = \frac{1}{4} \dot{M} v_w^2
= 1.9 \times 10^{34}
\left( \frac{\dot{M}}{3.0 \times 10^{-8} M_\odot yr^{-1}} \right)\\
\times \left( \frac{v_w}{2000\ km\ s^{-1}} \right) ^2\ ergs\ s^{-1},
\end{multline}

\noindent where $\dot{M}$ is the stellar mass loss rate and $v_w$ is the terminal wind speed. Comparing this and the previous equation we see that if, as proposed above, a substantial fraction of the wind energy is converted into gas thermal energy, then the heating can exceed the cooling by a large amount. Then the temperature of the gas in the mixing region can be raised well above the values in typical HII regions.

Specifically, the gas in the inner mixing region will be heated to a new steady state temperature of up to $10^6$ K. This is much like the immediate post-shock gas in the classic hydrodynamical models. In that case all the heating is assumed to occur within the thin shock (e.g., a thin Weibel growth region). This is in contrast to the picture suggested here of particle heating throughout a wider mixing region, though it is strongest within about one collisional dissipation length (e.g., from equation \eqref{eqb}) in the Weibel zone. 

In fact, the primary cooling process in the hot zone is probably outflow along the sides and subsequent adiabatic expansion. The outflowing energy rate is order $\rho A c^3$, where $\rho$ is the ambient mass density, $c$ is the sound speed, and $A$ is the outflow area. Equating this to the wind dissipation, e.g., about half or less of the wind luminosity of equation \eqref{eqh} above for the dissipation over one dissipation length, we can solve for the sound/expansion speed. Assuming again that $n \simeq 10$ cm$^{-3}$ and using the same dimensions for the emission region as above, we obtain a value of order 100 km s$^{-1}$ and a corresponding temperature of order 10$^5$ K. This temperature will decrease over each succeeding collisional dissipation length.

\subsubsection{Sputtering}

In the gas with the highest temperatures in the hot zone there is a moderately short grain sputtering timescale. With the gas and grain parameters assumed above, the sputtering timescale is of order a few times $10^5$ yrs, for temperatures of order $ 3 \times 10^5$ K. Grains traveling at more than 20 km s${-1}$ traverse a distance of more than 6 pc in a sputtering timescale, and so will flow out of the hot zone in about one destruction timescale. The sputtering rate falls rapidly as temperatures decrease, so there will be little sputtering outside the hottest layer. Thus, if we assume a temperature gradient from the inner, hottest parts of mixing regions to cooler, outer parts, where fast wind particles are depleted, then grains may penetrate most of the mixing region. Thus, the IR emission is associated with grain distress and destruction, via thermal spiking as well as collisional and stellar heating, and some sputtering. Grain depletion downstream may also help explain why the IR bow nebulae do not wrap around the star, even though high speed wind particles hit flowing ISM gas at essentially all azimuths. 

\subsubsection{Gas-grain interaction zone}

Ultimately, much of the wind particle energy will be available to help power the infrared emission. However, the net cross section of the dust grains to high speed protons is small; so their energy cannot be transferred directly to the dust. Rather, the stellar wind energy is dumped into the ambient particle population as discussed above. Collisions between these slower particles and grains can help power the IR emission. 

As a specific example, assume an ambient gas density of 10 cm$^{-3}$, a dust-to-gas mass density ratio of about 0.01 (outside any hot, sputtering zone), a grain radius of 0.1 $\mu$m, and a grain internal density of about 3.6 g cm$^{-3}$. Then the number density of the grains is about $6.6 \times 10^{-12}$ cm$^{-3}$ and the mean free path of a particle through the grain ensemble is, 

\begin{equation}
\label{eqc}
\lambda_{gr} = \left( n_{gr} \sigma_{gr} \right) ^{-1}
= 150 \left( \frac{0.1\ \mu m}{r_{gr}} \right) ^2
\left( \frac{10\ cm^{-3}}{n} \right)\ pc,
\end{equation}

\noindent where $n$ is the gas number density, $n_{gr}$ grain number density, and $\sigma_{gr}$ is the grain cross section. This estimate confirms that fast protons are very unlikely to hit a dust grain directly, unless they lose much of their energy (e.g., via electron scattering as described above). 

Secondary protons with a mean velocity of, for example, about 50-100 km s$^{-1}$, have a mean free path for mutual collisions of about 0.0042-0.034 pc or $(0.013-1.0) \times 10^{17}$ cm according to equation \eqref{eqb}. Their random walks will cover a pathlength of 150 pc after about 4400-35,000 scatterings. In a random walk this will yield a mean travel distance from the starting point of about (0.079-2.1) pc. so these slower, denser protons, located several dissipation lengths downstream from the hottest gas, have a good chance of hitting a grain before diffusing far from the mixing region. 

\subsection{Grain temperatures in the emission region}
\subsubsection{Thermal balance} 

It is usually assumed that the IR emitting grains in bow nebulae are heated by the photons from the central star; even late O-type stars are very luminous. However, in the parameter range considered here collisional heating in the mixing zone must also be considered. To estimate the photo-heating we begin by adopting the grain model of the previous subsection. Specifically, we assume grains of size 0.1 $\mu m$, adopt the same emission volume as above (i.e., assuming both a radius and thickness of the hemispheric emission region of 0.2 pc), and assume the same grain density, which yields a number of grains in the emitting region of about N = 9.9 $\times$ 10$^{41}\ n$. Then the total grain absorbing surface area to outgoing photons is,

\begin{equation}
\label{eqd}
N Q_{abs} \left( \pi r_{gr}^2 \right) = 3.1 \times 10^{32}
n Q_{abs}   \left( \frac{r_{gr}}{0.1\ \mu m} \right) ^2\  cm^2,
\end{equation}

\noindent where $n$ is the gas number density, and $Q_{abs}$ is the photo-absorption efficiency. Dividing this quantity by the hemispheric surface area at the standoff radius gives the fraction of stellar photons intercepted by the mixing zone grains. It is,

\begin{equation}
\label{eqe}
f_{int} = 0.00013
n Q_{abs}   \left( \frac{r_{gr}}{0.1\ \mu m} \right) ^2.
\end{equation}

The corresponding intercepted luminosity is obtained by multiplying this by the stellar luminosity, e.g., of order 10$^5$ L$_\odot$, we obtain,

\begin{multline}
\label{eqf}
L_{int} 
= 4.9 \times 10^{34}
n Q_{abs}   \left( \frac{r_{gr}}{0.1\ \mu m} \right) ^2
\left( \frac{0.2\ pc}{R_o} \right) ^2\\
\times
\left( \frac{L}{10^5\ L_\odot} \right)\ ergs\ s^{-1}.
\end{multline}

Next, we assume that this intercepted luminosity equals the thermal grain emission. This is about $N(4\pi r_{gr}^2) \epsilon \sigma T^4$, where $\epsilon$ is a radiative efficiency and $\sigma$ is the radiation constant. Equating this expression to $L_{int}$ from equation \eqref{eqf}, using equation \eqref{eqd}, and solving for $T$ we obtain,

\begin{equation}
\label{eqg}
T_{gr,photo} = 29  \left( \frac{Q_{abs}}{\epsilon} \right) ^{1/4}
\left( \frac{0.2\ pc}{R_o} \right) ^{1/2}\ K.
\end{equation}

\noindent Note that the direct dependences on grain size and number density cancel. The efficiency $\epsilon$ is somewhat smaller than $Q_{abs}$ at the relevant temperatures and densities (see \citealt{draine11}, Ch. 24), but that will only increase the temperature estimate by a factor of about a couple. Observations estimate the grain temperatures at $ >$ 100 K in many systems \citep{kobulnicky17}, so this is  an indication that photo-heating is not sufficient, especially for late-O and B stars (see discussion in the following section).

We can estimate the grain heating due to the stellar wind similarly. That is, assume that the kinetic luminosity of the stellar wind is transferred to grain heating, via heating the mixing region as above, and that this is balanced by grain cooling. The wind luminosity is given by equation \eqref{eqh}, with the addition of an efficiency factor $\chi$ for the energy transfer to grains. Equating this expression to the IR luminosity we obtain an estimate of the grain temperature resulting from the particle impact heating,

\begin{multline}
\label{eqi}
T_{gr,imp} = 68 \chi
\left( \frac{\dot{M}}{3.0 \times 10^{-8} M_\odot yr^{-1}} \right) ^{1/4}
\left( \frac{0.0013}{\epsilon}\right) ^{1/4}\\
\times  \left( \frac{v_w}{2000\ km\ s^{-1}} \right) ^{1/2}
\left( \frac{10\ cm^{-3}}{n} \right) ^{1/4}
 \left( \frac{r_{gr}}{0.1\ \mu m} \right) ^{1/2}K,
\end{multline}

\noindent where we have adopted a value of $\epsilon = 0.0013$ at about 100 K (see \citealt{draine11}, Ch. 24). The grain size is likely somewhat smaller in the emission region; even a factor of a few decrease would push the grain temperature above 100 K, for $\chi \le 1$. We conclude that over much of the relevant parameter range particle heating can be comparable to photo-heating of the grains, a topic we will explore further in the next section.

\section{Powering the Emission}
\subsection{Stellar heating deficiency}

\citet{kobulnicky17} noted that the colour temperatures of bow nebulae are 1.1-3 times higher than expected from stellar heating of the grains and postulated an additional heating source (see their Fig. 10). In this subsection we explore several approaches to the comparison of heating processes. These approaches are based on a comparison between stellar and IR nebular luminosities provided by \citet{kobulnicky17} for $19$ stars (see their Table 5), and updated with about 70 stars in \citet{kobulnicky19}.  They find the star to nebular luminosity ratios range from about $1$ to $22,422$ with a median of about $300-400$, though there are significant uncertainties in these ratios. Nonetheless, given that less than half of the starlight is directed towards the bow nebula, and that only a small fraction is intercepted (see equation \eqref{eqe}) for a low-to-moderate density ambient medium, it appears that in most cases the direct stellar energy input falls far short of what is needed to power the nebulae. \citet{acreman16} addressed this question of powering the nebular by stellar emission in some detail. They find such powering plausible, but they assume a stellar spectral type and luminosity equal to that of the brightest stars in the \citet{kobulnicky19} tables. For the star with the highest stellar to IR nebula luminosity ratio in the Kobulnicky et al. tables, an estimate based on equation \eqref{eqe} with $n = 1.0$ cm$^{-3}$ and the assumption that the nebula intercepts a full hemisphere of starlight, yields an intercepted fraction of about 0.35 of the IR luminosity. This suggests that even systems with high stellar to nebular luminosity ratios might not be fully powered by stellar heating. On the other hand, in an ambient medium with $n = 1000$ cm$^{-3}$ (and assuming the same dust-to-gas ratio), starlight could power the IR emission in most of Kobulnicky  et al.'s systems. In such cases, however, equation \eqref{eqe} suggests that a large fraction of the stellar luminosity is intercepted, which is generally not observed, except in the Bubble Nebula and RCW 120, which have a couple of magnitudes of visual extinction (see discussion in the following section). 

\subsubsection{Stellar versus wind heating}

 As a second way of looking at the heating processes, we again use the data on bow nebulae and their parent stars assembled by \citet{kobulnicky19}  (see their Tables 1-3, and also \citealt{kobulnicky17} ). In Fig. 2 we plot their adopted standoff radii $R_o$ versus terminal wind velocities (the latter were adopted from \citealt{mokiem07} and some of these are model based) for 69 stars. Also shown on this plot is a solid curve based on equating the wind luminosity (equation \eqref{eqh}) to the intercepted stellar luminosity (equation \eqref{eqf}), and solving for the wind velocity as a function of $R_o$. The mass loss factor in eq. 10 is replaced using the equality,

\begin{equation}
\label{eqj}
\dot{M} = 4 \pi R_o^2 \rho_w v_w 
= 4 \pi R_o^2 \rho v_{rel} \frac{v_{rel}}{v_w} ,
\end{equation}

\noindent where it is assumed that at the standoff radius, the momentum flux of the wind equals that of ambient medium, $\rho_w v_w  = \rho v_{rel}$. The subscript {\it w} refers to wind quantities, and $\rho$ is the interstellar mass density near the radius $R_o$. 

\begin{figure}
\centerline{
\includegraphics[scale=0.38]{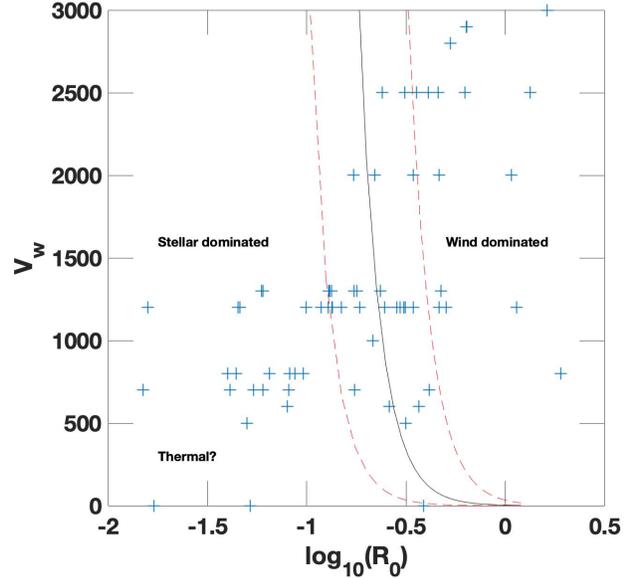}}
\caption{Wind speed ($v_w$) versus the logarithm of the standoff radius ($R_o$). The plus signs represent the stars tabulated in \citet{kobulnicky19}. The solid curve shows $v_w(R_o)$ when the stellar heating and maximal wind heating are equal. The dashed curves show the effect of increasing the luminosity ratio by a factor of 10 or 0.1. See text for details.}
\label{fig:f2}
\end{figure}

Thus, the solid curve illustrates the boundary between regions where the maximal mechanical luminosity of the stellar wind exceeds the stellar luminosity intercepted by the grains, or vice versa. Wind dominated regions lie above and to the right of the curve. The dashed curves have wind velocities ten times greater or ten times lesser at a given value of $R_o$ than the solid curve. They are included to highlight the fact that, because of the steepness of the solid curve, substantial changes in the model wind speeds (and consequently the mass loss) would not greatly change the result that there are a number of systems on each side of the curve. 

If only a fraction of the wind luminosity goes into grain heating the solid curve will shift to the right. However, the right hand dashed curve, indicates again that even a substantial change in the available energy fraction does not change the overall conclusion. Wind luminosity is likely to contribute significantly to grain heating in systems with $R_o > 0.25$ pc., and direct stellar heating probably dominates in most cases with $R_o < 0.15$. 

Another possible heating source is the thermal energy in the ambient HII region gas. We can estimate this as \citep{draine11},

\begin{multline}
\label{eqk}
L_{th} = 3.4 \times 10^{31} \alpha_T
\left( \frac{n}{100\ cm^{-3}} \right) 
\left( \frac{v_{th}}{200\ km\ s^{-1}} \right)\\ 
\times \left( \frac{a}{0.1\ \mu m} \right)^2  
\left( \frac{T_g}{20000\ K} \right)^{3/2}\ ergs\ s^{-1}.
\end{multline}

\noindent We can assume that the grain heating efficiency $\alpha_T$ is unity to maximize this estimate. This estimate suggests that this source of heating is likely to be small unless the ambient density is quite high. In that case, $R_o$ will be small and wind heating will be unimportant, unless there is an unusually high wind velocity. Thus, we have placed the possible thermal heating region in the lower left of Fig. 2. 

\subsubsection{Combining heating sources}

\begin{figure}
\centerline{
\includegraphics[scale=0.38]{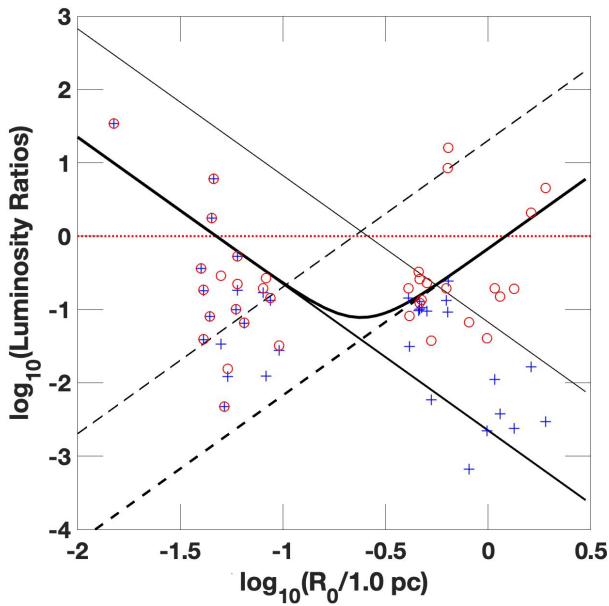}}
\caption{Logarithm of various observed and analytic luminosity ratios relevant to grain heating versus the logarithm of the standoff radius ($R_o$). Solid lines give estimates of stellar heating rates normalized to those of an OV8 star in the cases of an ambient medium density of $n_a = 10$ cm$^{-3}$ (thick line), and $n_a = 300$ cm$^{-3}$ (thin line). The dashed lines gives corresponding estimates of the maximal stellar wind heating in the same two ambient medium densities and with the same normalization. See text for details. The thick curve is the sum of the two thick lines. The plus signs show estimates of the ratio of intercepted stellar luminosity to observed infrared bow nebula luminosity for selected systems from \citealt{kobulnicky19} (see text for details). The circles show the sum of this ratio and the estimated ratio of maximal wind heating to the observed IR luminosity. The horizontal line shows when these ratios equal unity, i.e., when the heating sources equal the observed IR luminosity.}
\label{fig:f3}
\end{figure}

In the preceding paragraphs, we have considered the relative roles of stellar and wind luminosities. The question of whether either or both provide sufficient energy input to power the observed IR nebulae remains. We can address this question from another angle with the aid of Fig. 3. This figure shows various analytic and observed luminosity ratios versus standoff radii $R_o$. For example, the thick, downward trending solid line shows the stellar luminosity intercepted by grains according to equation \eqref{eqf}, assuming $Q_{int} = 0.1$, and $n = 10$ cm$^{-3}$, and then divided by a normalization factor. This factor is the luminosity of a typical OV8 star (taken as $3.45 \times 10^{38}$ ergs s$^{-1}$) divided by 400, which is a typical ratio of stellar to IR nebula luminosities. Thus, the normalization factor is a representative bow nebular luminosity, so the line shows an estimate of the intercepted stellar luminosity compared to  a typical IR nebula luminosity. The thin down-trending line above it is the same, except with a higher ambient density of $n = 300$ cm$^{-3}$. The up trending lines are the analogous luminosities for maximal wind particle heating according to equation \eqref{eqh} with $n = 10$ cm$^{-3}$ (thick dashed line) and $n = 300$ cm$^{-3}$ (thin dashed line), and the same IR nebula normalization. As in equation \eqref{eqh} these lines assume complete conversion of wind kinetic energy (in one hemisphere) into grain heating, and so, are maximal estimates. 

The sum of these two energy ratios is shown in the figure by the thick curve. It is clear that stellar heating dominates at low values of $R_o$, and wind heating at high values of $R_o$, unless the conversion factor $\chi$ of wind energy into grain heating is very low. The two processes are comparable only in a narrow range of $R_o$ values. The dotted horizontal line shows where these ratios equal unity. I.e., where the input luminosities equal the luminosity of the normalization factor, or a typical bow nebula luminosity. It appears harder to form bow nebulae of intermediate size ($log_{10}(R_o) \simeq -0.6$), unless there is a relatively high density ambient medium.

The plus signs in Fig. 3 show the ratio of the intercepted stellar luminosity to observed IR bow nebula luminosity for 16 stars with the highest values of $R_o$ ($log_{10}(R_o) > -0.4$) and 21 with the lowest values ($log_{10}(R_o) < -1.0$) from the sample of Kobulnicky, et al. (2019). Other stars in the sample were omitted for clarity in the figure. Specifically, the intercepted luminosity was computed from equation \eqref{eqf} with $Q_{int} = 0.1$, but using the particular values of the stellar luminosity, the ambient density and $R_o$ for each star obtained from the tables of Kobulnicky, et al. (2019). The infrared nebula luminosities were obtained from the same source. 

The circles in Fig. 3 are the derived by adding the total wind luminosity (from equation \eqref{eqh}) to the intercepted stellar luminosity and dividing by the IR nebula luminosity. The values of the mass loss and wind velocity used to compute the wind luminosity were again obtained from the tables of Kobulnicky, et al. (2019). We emphasize again that there is substantial, and practically unquantifiable, uncertainty in these quantities, as well as some of those used to compute the intercepted luminosity. Granted that, this plot suggests a number of interesting results. 

Firstly, as expected, the (maximal) wind luminosity dominates in most cases at high values of $R_o$. Alternately, the intercepted stellar luminosity fails significantly, often by orders of magnitude, to account for the IR emission for most of the high $R_o$ stars. Secondly, the opposite is true for the stars with low values of $R_o$. Only in a few cases, in this limit, does the wind luminosity add significantly to the intercepted stellar luminosity. (Many of these stars are B or late O types, with relatively low mass loss rates.) Thirdly, most of the circles are well below the dotted line of equality, meaning that the estimated sum of these two luminosities does not provide enough power to account for the IR emission. The combined data uncertainties may be at least a factor of a few, putting many of these values within range of equality. We will consider other factors below. Fourthly, a few of the circles are well above the line of equality. For such cases at high $R_o$ the wind heating of the grains is likely over-estimated, i.e., not all of the wind luminosity may be converted into IR emission. This is also true generally, and a correction factor would pull many systems farther below the line of equality. For the few points with low values of $R_o$, in the upper left of the figure, the wind luminosity does not contribute significantly, so another explanation is needed. 

Given the possible inadequacy of the stellar wind heating in many cases we consider another factor in equation \eqref{eqf}, the average grain size. Assuming still the constancy of the the total grain mass to gas mass ratio, but supposing that the grains are broken down (i.e., by collisions with energetic particles, see e.g., \citealt{pavlyuchenkov13}) to an average size of a factor of a few less than assumed in that equation, then all of the points in Fig. 3 would be raised by about an order of magnitude. This is because of the large increase in total surface area and intercepted starlight with the smaller grains. This would also increase the grain cross section for intercepting energetic particles, which might result in the destruction of more small particles. Small particles could also be destroyed by thermal spiking, which may contribute significantly to the IR emission. Thus, it seems unlikely that the emitting grain population could be dominated by those with sizes much smaller than previously assumed, or that the whole luminosity problem could be solved with such a population. Moreover, dominance by small grain populations can be constrained by upper limits on the near infrared emission of the nebulae. 

Similarly, part of the luminosity problem could be solved if the bow nebulae acquire a higher than average grain-to-gas mass, i.e., if grains accumulate in the nebulae. The above mean free path estimates suggest that the grains are not too tightly coupled to the gas in these conditions. Thus, in the outer mixing region, as the ISM wind flow is decelerated and turned aside, grains may flow forward into the region.  This extra inflow could replenish any losses resulting from sputtering and outflow. According to the classification of \citet{henney19a}, cases with stellar and ISM parameters like those considered above are in the conventional `bow wind' group, not the thick `dust wave' group. On the other hand, outflow, and high temperature sputtering in the inner part of the mixing region argue against grain buildup. If such a grain buildup did occur, the grain opacity would increase, and it may be detectable as absorption of background sources in the optical or near-infrared bands. 

In conclusion, it seems likely that intercepted stellar luminosity is insufficient to power the infrared emission in most observed bow nebulae given a typical mass and size distribution of the dust grains. The reprocessing of the stellar wind luminosity into grain heating can help considerably to explain this luminosity deficit in some cases. The combination of all relevant processes including: the interception of stellar luminosity, the reprocessing of wind luminosity, a modest reduction of the mean grain size, and thermal spiking in small grains, could make up the deficit in almost all systems. Variations in these processes depend on wind properties and the velocity of the star through the ambient medium. For example, consider the four circles in the upper right part of Fig. 3, all above the unity line. Two of these stars are supergiants, and the other two are O5V stars, among the earliest spectral types hosting bow nebulae. All of them have very high inferred mass loss rates. Three of the four have velocities relative to the ambient medium of more than 30 km s$^{-1}$. If these factors result in extreme grain breakdown, or faster flow through the bow region preventing grain buildup, then the circles could be brought down to the unity line.

\section{Comparison to Observed Systems}

As noted in the introduction, a few nearby bow nebula systems have been observed in multiple wavebands, and on a large enough scale to possibly detect all the main components. The bow nebulae were discovered and are most prominent in the mid-infrared, and are generally expected to be found on much smaller scales than the ionization shock bubble, at least if the surrounding ISM is of relatively low density. The ionization bubbles will likely be intrinsically faint, but potentially observable in the near infrared, optical and ultraviolet.

\subsection{$\zeta$ Ophiucus and RCW 120}

Two of the best studied bow nebulae are $\zeta$ Ophiucus and RCW 120. Although noted in many papers, the former has not been specifically modeled. The latter was modeled by \citet{mackey15}. Fig. 1 of \citet{mackey15} shows Spitzer Space Telescope $8$ and $24$ micron images of the system. The bow nebula is only prominent in the latter, and barely visible in the former band, and presumably, this faintness is also true in shorter wavelength observations. We expect that strong shocks in the vicinity of the bow nebula would have significant optical and higher energy emission, while a lower temperature Weibel instability zone may not. For example, the runaway star bow shock models of \citet{meyer16} predict substantial optical line emissions, though with the caveats that the models discussed in the most detail had relatively fast space velocities ($40$ and $70$ km s$^{-1}$) compared to those considered above. 

The paper of \citet{mackey15} is notable for using (two-dimensional) radiation-hydrodynamic simulations to model both the HII region discontinuity and the (presumed) strong bow shock within it. It is also unusual in using low stellar space velocities, i.e., $4 - 16$ km s$^{-1}$, which make the ionization front an almost negligible discontinuity. A fairly high ambient ISM density of $3000$ cm$^{-3}$ was assumed, the star may be in its natal molecular cloud. However, this density is perhaps a bit too high to fit the surroundings of RCW 120 according to the authors. With these parameters the authors were able to reproduce general features of the RCW 120, including the observation that the radius of the HII region bubble is only a few times larger than the radius of the bow nebula. This is a natural consequence of the high ambient density. The shock in the Mackey et al. models should generate substantial optical and UV line emission. They also predict significant X-ray emission, but argue that much of it may be absorbed within the HII region. They note substantial (hydrodynamic) mixing in the contact regions of the models and point out that we might expect the line emission to be spread across the larger mixing region, where thermal conduction may also transport energy away. 

The authors favor their lowest velocity model to best fit the RCW 120 morphology. However, it might be hard to account for the $8$ micron bubble emissions in such a model, though no quantitative comparisons were made. Moreover, the greatest morphological incongruity in their higher velocity models (e.g., $16$ km s$^{-1}$) is in the bow shock. A diffusive instability front, as described above, would not show the same narrow morphology, and thus, might be more consistent with the observed morphology, as well as the emission characteristics. 

\subsection{Observed outer shells}

A number of objects in the bow nebula catalog of \citet{kobulnicky16} show both a bow nebula, visible primarily at $24$ microns (in Spitzer or 22 microns in Wise data), and an outer shell, which is generally visible in the Spitzer $8$ micron band. These include catalog objects:  42, 81,142, 214, 231, 309, and 342, and up to a few dozen others with less regular morphologies. Objects 42, 231 and 342, in particular, look much like RCW 120. In all of these objects the outer shell is at a radius of only a few times (or less) that of its IR bow nebula. Thus, if these outer shells are in fact shock-ionization fronts, then a relatively dense surrounding medium is required, probably the stellar natal molecular cloud in most cases. Estimates for the ambient density of these objects are not available in \citet{kobulnicky19}, except for object 342, which has a listed value of 795 cm$^{-3}$. 

These outer shells have been explained as evaporation nebulae, where dense clumps in the cloud are heated by the stellar UV photons (e.g., \citealt{kobulnicky16}). They may simultaneously be dust waves as as described by \citet{ochsendorf14} for $\sigma$ Orionis AB or \citet{henney19a}. Aside from the objects listed above, a number of objects in the Kobulnicky et al. catalog have quite irregular $8$ micron emission regions nearby. We might expect such irregularity if the star is heating and evaporating the nearest clumps in a turbulent molecular cloud. Outer bow-shaped emission regions would seem much less likely in such environments, and so, they may more likely be ionization fronts (and/or dust waves), though more evidence is needed to test that. 

Many objects in the \citet{kobulnicky16} catalog have no visible outer shell, only the IR bow nebula. This is not surprising since most of the published images do not extend more than a few times the radius of the bow nebula, so in a low-to-moderate density ambient medium even the nearest part of an ionization front would be outside the image. Unless the relative velocity was high, the outer front would also be quite faint in most wavebands. Clearly, studies with a wider field of view in a variety of wavebands would be helpful in resolving the nature of the smaller-scale, IR bow nebulae.

\subsection{The Bubble Nebula NGC 7635}

The Bubble Nebula may be another nearby example of an infrared bow nebula. The recent study of the object by \citet{green19} summarizes measured properties, including: stellar type and temperature (O6.5, 37,500 K), mass loss rate, wind velocity (2500 km s$^{-1}$), and peculiar velocity (about 28 km s$^{-1}$). This is considerably more information than we have for most systems. It is apparently not in the \citet{kobulnicky18} catalog, and its Spitzer 24 micron image does not have a completely clear bow shape. Both high resolution Hubble Heritage optical image (https://hubblesite.org/image/3725/gallery) and a large-scale emission line image from Mt. Wilson are available \citep{jurasevich10} (also Astronomy Picture of the Day, 2010 Sept. 2). The latter shows a large fragmented shell-like structure, with a diameter a few times that of the usual `bubble'. The former suggests that the brightest emission region in the bubble is either an engulfed cloud fragment, or an illuminated cloudlet just outside it. \citet{green19} estimate the bubble diameter to be about 2.5 pc., though the asymmetrically located star comes within about 0.3-0.4 pc of the bubble edge on the forward side. The outer shell diameter is roughly 2.5 times that of the bubble, or about 6 pc.

The \citet{green19} study includes new two-dimensional hydrodynamic models and detailed comparisons to the morphology of the nebula. As in previous studies, the bow nebula emissions were modeled as those of the cooling regions behind a conventional bow shock. A bubble ring is produced at particular viewing angles of the models. The models also show a dense stellar wind 'bubble', though in some snapshots the highest density contour is deformed far from a spherical shape. The models produce a great deal of downstream turbulence. The turbulent vortices do not well resemble the filaments seen in high resolution images. There was no attempt to reproduce the emission clumps (inside and outside the bubble) observed in various wavebands. 

The models produce a hot, million degree, layer, which in turn, produces substantial X-ray emission, which is not observed. However, the authors point out that the nebula is observed through 2 magnitudes of optical extinction, which could absorb some 99\% of the X-rays. The models also produce H$\alpha$ and mid-IR shells like the observed ones in size and general morphology, at the appropriate viewing angles. The models don't seem to produce the filamentary structure observed. The models also do not generally produce the large-scale outer shell seen in the Mt. Wilson image. 

The large-scale shell might be explained as a weak ionization shock. \citet{green19} estimate an ambient density of about 100 cm $^{-3}$, in which case the classical Stromgren radius would be about 3 pc \citep{osterbrock89}, comparable to the large-scale shell. This shell appears filamentary at all locations, and quite broken up in the forward direction. The emission filaments and polyps seen on the high resolution images in that direction suggest that the star is moving into a clumpy, filamentary, and high density cloud in that direction. The cloudlet projected onto the bubble reinforces this impression. If this interpretation of the large-scale shell as near the ionization edge is correct, then probably all the bright emission structures are excited by ionizing radiation. 

Using the parameters listed in \citet{green19} we can estimate the standoff radius as about 1.1 pc. It is very possible that the ISM density is greater than the assumed value of 100 cm$^{-3}$ in the forward direction, so accounting the shorter star-to-bubble edge distance in that direction. Similarly, the ISM density may be less in the downstream direction, accounting for the greater separation. Thus, in accord with the assumption of the hydrodynamic models, the bubble surface may be the primary interface between wind and ISM. The models show a sharp change in density and temperature at this surface, though the optical and IR maps derived from them show somewhat broader surface, due to projected emissions from different locations. However, they do not show the same filamentary structure visible around the bubble in the observations, and especially not  the wide emission region in the forward direction. 

This structure is suggestive of a broader diffusive mixing region. If the bubble is a diffusive Weibel zone, then we do not need projection effects to account for it. We simply observe (a thicker shell than that of a shock) through a larger column density at the edges, regardless of projection angle.  

The clumpiness and complex structure visible in the observations, especially in the forward direction, suggests that ionizing (and sub-ionizing UV) radiation may leak out of the nebula inhomogeneously, and power the more distant and faint emission regions visible especially on the Mt. Wilson and other telescope images (several available on the Astronomy Picture of the Day website). 

More generally, it is clear even in this well studied case, as well as the earlier examples, that more systematic, multi-waveband and multi-scale observational studies are needed to elucidate these objects. 

\section{Fast runaway stars}
\label{runners}

The phenomena discussed in the previous sections live in a restricted range of stellar velocities through the ISM. For stars with very low peculiar velocities, the standoff radius will be very large, the hemispheric wind luminosity will be shared with a much larger number of grains, so the mean grain temperature and emission will be much lower. Dust accumulation and radiation pressure may be greater, and a `dust wave' (\citealt{henney19a}, also see \citealt{henney19b, henney19c, henney19d}) may form, either within or outside the beam dump.

The situation is different for high speed stars. The term 'runaway' star has various definitions in the literature, including velocities as low as those considered in the previous section. Here we consider hypervelocity stars with velocities $>$ 50 km s$^{-1}$ relative to the local ISM. \citet{peri12, peri15} find that in a sample of runaway stars (including some with velocities of only tens of km s$^{-1}$) only about 5\% have detectable IR emission. Many runaway stars are no longer in the thin disc of the Galaxy, so the ambient gas may be of very low density, preventing much emission. However, there are other reasons for the lack of IR emission.

The ISM ram pressure of fast, hypervelocity stars (viewed in the stellar frame) is much greater than the thermal pressure of the photo-ionized and photo-heated gas at the ionization edge, and also greater than the typical magnetic pressure in the ISM. When these fast flows hit that edge, the gas will be ionized and photo-heated, but the flow will remain supersonic. 

The standoff radius will be significantly closer to the star compared to lower velocity cases. An ISM shock will form as in the bow shock models referenced above, and this will compress and heat the turbulent inner mixing zone. Grains that penetrate into the shocked region may melt or sputter due to both the more intense photo-flux and the shock heated gas. Thus, strong IR bow nebulae seems unlikely in hypervelocity stars. 

\section{Summary \& Conclusions}

In the previous sections we have seen how at moderate relative velocities (a few tens of km s$^{-1}$), ionization, photo-heating, and gyro-motions around ambient fields will initially thermalize the ISM gas at a temperature corresponding to a sound speed of about a few times less than the relative velocity. Generally, a weak, R-type ionization front will form slightly interior to the classical Str\"{o}mgren edge, and well outside the standoff radius between the ISM flow and the stellar wind. Behind this front ISM material flows transsonically. This is a special case of a textbook result. 

When stellar wind particles meet the ISM flow a mixing region will form mediated by the Weibel instability. Stellar wind protons have a very long mean free path for scattering off ambient protons in a gas of moderate density (e.g., 1-10 cm$^{-1}$), so wind ions will mix with the ISM gas within the filamentary Weibel turbulence region and beyond its dissipation zone. Ultimately, the ions distribute their energy in a region near the momentum balance surface, though with a substantial extent in radius. Specifically, the thickness of the IR emission region is predicted to be of order the wind proton diffusion length discussed in Sec. 2.2.1. The inner part of this mixing region is predicted to be heated to temperatures substantially above that of the surrounding HII region (though not as high as in hydrodynamic shock models), since the strong ionization suppresses cooling and line emission. The temperature gradient is also expected to be shallower than in hydrodynamic shock models.

As noted above, \citet{kobulnicky17} found that the colour temperatures of the bow nebulae are 1.1-3 times higher than expected from stellar heating alone and suggested the existence of an additional heating source. The processes described above can provide most of the energy for the IR emissions. That is, while direct stellar heating, or heating from wind particles, are generally inadequate individually, together these processes can account for the heating in many systems. In other systems grain buildup in the standoff region or small mean grain size are required to increase the total grain cross section to the primary heating processes. As discussed in the previous section, the theory also explains why the nebular emission only occurs in a limited range of relative velocity. 

Another prediction of this theory is that since it takes some time, and some travel distance, to deposit stellar proton energies, any stellar and wind variabilities should be smoothed in the IR emission. This may not be true in higher energy emissions, like H$\alpha$, originating in sharper bow shocks in hypervelocity stars. Detailed particle+hydrodynamic simulations will be needed to test this theory and its predictions in more detail and calculate the IR emissions.

\section*{Acknowledgments}
I am grateful to Chip Kobulnicky for making me aware of the bow nebula phenomenon and to Chip and Rico Ignace for helpful input. I am grateful to the reviewers for contributing significant improvements to the original version. 

\bibliographystyle{mn2e}

\bsp
\label{lastpage}
\end{document}